\documentclass[11pt]{article}
\usepackage{graphicx}
\usepackage{color}
\usepackage{amsmath}
\usepackage{amssymb}
\usepackage{mathrsfs}
\usepackage{textcomp}
\usepackage{braket}
\usepackage{pifont}
\usepackage{bbding}
\usepackage{fancybox}
\usepackage{array}
\usepackage{longtable}
\usepackage{latexsym}
\usepackage{varioref}
\usepackage{xspace}
\usepackage{makeidx}
\usepackage{verbatim}
\usepackage{graphicx}
\usepackage{tabularx}
\usepackage{fancyhdr}
\usepackage{titletoc}

\newtheorem{theorem}{\indent Theorem}

\newcommand{\xLongLeftRightArrow}[2][]{%
  \ext@arrow 0055{\LongLeftRightArrowfill@}{#1}{#2}}
\def\LongLeftRightArrowfill@{%
 \arrowfill@\Leftarrow\Relbar\Rightarrow}

\title{Commutative-like Encryption: A New Characterization of ElGamal}
{\author{\small DAI Wei $\footnote{DAI Wei is at department of
computer science and technology, Tsinghua University. }$\\
}}

\date{}
\begin{document}
\maketitle
\begin{abstract}
Commutative encryption is a useful but rather strict notion in
cryptography. In this paper, we define a loose variation of
commutative encryption-commutative-like encryption and give an
example: the generalization of ElGamal scheme. The application of
the new variation is also discussed.
\end{abstract}

\section{Introduction}
Informally, a commutative encryption is a pair of encryption
functions $f$ and $g$ such that $f(g(v))=g(f(v))$. Commutative
encryption is extremely useful in modern cryptography since many
protocols rely on the existence of commutative
encryption$\cite{commutativeA, commutativeB, commutativeC,
commutativeD}$. However, few encryption schemes are known to be
commutative. In this paper, we introduce a loose notion of
``commutative-like encryption'' and propose a primitive: the
generalization of ElGamal. First introduced by
ElGamal$\cite{ElGamal}$, the ElGamal encryption is one of the most
famous public key encryption schemes and has various
applications$\cite{elgamal1, elgamal2, elgamal3}$. Based on
ElGamal encryption, this new characterization shares most
advantages of commutative encryption and ElGamal while the
definition itself is not as strict as commutative encryption.

\section{Preliminaries}
We first describe some relevant definitions that would be used in
the paper.

\subsection{Commutative encryption}

Our definition of commutative encryption below is similar to the
constructions used in $\cite{commutative, direction}$ and others.
As showed above, a commutative encryption is a pair of encryption
functions $f$ and $g$ such that $f(g(v))=g(f(v))$. Thus by using
the combination $f(g(v))$ to encrypt $v$, we can ensure that
$\mathcal{R}$ cannot compute the encryption of a value without the
help of $\mathcal{S}$. In addition, even though the encryption is
a combination of two functions, each party can apply their
function first and still get the same result.

DEFINITION 1\ \ (Indistinguishability). \emph{Let
$\Omega_k\subseteq\{0, 1\}^k$ be a finite domain of $k$-bit
numbers. Let $\mathcal{D}_1=\mathcal{D}_1(\Omega_k)$ and
$\mathcal{D}_2=\mathcal{D}_2(\Omega_k)$ be distributions over
$\Omega_k$. Let $A_k(x)$ be an algorithm that, given $x\in
\Omega_k$, returns either true or false. We define distribution
$D_1$ of random variable $x\in \Omega_k$ to be computationally
indistinguishable from distribution $D_2$ if for any family of
polynomial-step (w.r.t. $k$) algorithms $A_k(x)$, any positive
polynomial $p(\cdot)$, and all sufficiently large $k$,
$$|\mathrm{Pr}[A_k(x)|x\sim D_1]-\mathrm{Pr}[A_k(x)|x\sim D_2]|<\frac{1}{p(k)}$$
where $x\sim D$ denotes that $x$ is distributed according to $D$,
and $\mathrm{Pr}[A_k(x)]$ is the probability that $A_k(x)$ returns
true.}

Throughout this paper, we will use ``indistinguishable'' as
shorthand for``computationally indistinguishable''.

DEFINITION 2\ \ (Commutative Encryption). \emph{A commutative
encryption $\mathcal{F}$ is a computable (in polynomial time)
function $f$ : Key\ $\mathcal{F}\times$\ Dom\
$\mathcal{F}\rightarrow$\ Dom\ $\mathcal{F}$, defined on finite
computable domains, that satisfies all properties listed below. We
denote $f_e(x)=f(e, x)$, and use ``$\in_r$'' to mean ``is chosen
uniformly at random from''.}

\emph{ 1. Commutative. For all $e, e'\in$\ Key\ $\mathcal{F}$ we
have
$$f_e\circ f_{e'}= f_{e'}\circ f_e$$}

\emph{2. Each $f_e$: Dom\ $\mathcal{F}\rightarrow$\ Dom\
$\mathcal{F}$ is a bijection.}

\emph{3. The inverse $f^{-1}_e$ is also computable in
polynomial-time given $e$.}

\emph{4. The distribution of $\langle x, f_e(x), y, f_e(y)\rangle$
is indistinguishable from the distribution of $\langle x, f_e(x),
y, z\rangle$, where $x, y, z\in_r$\ Dom\ $\mathcal{F}$ and
$e\in_r$ \ Key\ $\mathcal{F}$.}

\subsection{ElGamal encryption}
We define the ElGamal public-key encryption scheme. The ElGamal
encryption scheme is based on the Diffie-Hellman assumption and it
is a probabilistic encryption scheme, i.e., a specific message has
many-exponential in the security parameter-possible encryptions.
Formally,

DEFINITION 3\ \ (ElGamal Public-Key Encryption
Scheme$\cite{ElGamal, elgamal security}$) \emph{The} ElGamal
public key encryption scheme \emph{is defined by a triplet $(G, E,
D)$ of probabilistic polynomial time algorithms, with the
following properties:}

\begin{itemize}
\item \emph{The system setup algorithm, $\mathcal{S}$, on input
$1^n$, where $n$ is the security parameter, outputs the system
parameters $(P, Q, g)$, where $(P, Q, g)$ is an instance of the
DLP collection, i.e., $P$ is a uniformly chosen prime of length
$P=n+\delta$ for a specified constant $\delta$, and $g$ is a
uniformly chosen generator of the subgroup $G_Q$ of prime order
$Q$ of $Z_P^*$, where $Q=(P-1)/\gamma$ is prime and $\gamma$ is a
specified small integer.}
\end{itemize}

\begin{itemize}
\item \emph{The key generating algorithm, $G$, on input $(P, Q,
g)$, outputs a public key, $e=(P, Q, g, y)$, and a private key,
$d= (P, Q, g, x)$, where $x\in_r Z_Q$, and $y\equiv g^x \bmod\
P$.}
\end{itemize}

\begin{itemize}
\item \emph{The encryption algorithm, $E$, on input $(P, Q, g, y)$
and a message $m\in G_Q$, uniformly selects an element $k\in_rZ_Q$
and outputs
$$E((P, Q, g, y), m) = (g^k (\bmod\ P), my^k (\bmod\ P))$$}
\end{itemize}

\begin{itemize}
\item \emph{The decryption algorithm, $D$, on input $(P, Q, g, x)$
and a ciphertext $(y_1, y_2)$, outputs $$D((P, g, x), (y_1,
y_2))=y_2(y_1^x)^{-1} (\bmod\ P)$$}
\end{itemize}

\section{Re-encryption}

In this section, we present a re-encryption algorithm of ElGamal.
Unlike most other schemes, using ElGamal encryption we obtain
ciphertext $(y_1, y_2)$, in this re-encryption algorithm, we need
not to encrypt $y_1$ and $y_2$ respectively, details follow (to
simplify the description, we still use the terms defined in the
previous section):

\begin{itemize}
\item \emph{To encrypt the plaintext $m$ (i.e., the ``first''
encryption step), we use the ElGamal scheme:}

\begin{itemize}
\item \emph{Key generation: Let $x_A$ be the element uniformly
chosen from $Z_Q$, and $y_A\equiv g^{x_A}\bmod\  P$.}

\item \emph{Encryption: On input $(P, Q, g, y_A)$ and a message
(plaintext) $m\in G_Q$, uniformly selects an element $k_A\in_rZ_Q$
and outputs
$$E((P, Q, g, y_A), m) = (g^{k_A} (\bmod\ P), m{y_A}^{k_A} (\bmod\ P))$$}
\end{itemize}

\item \emph{To re-encrypt the plaintext $(y_1, y_2)=(g^{k_A}
(\bmod\ P), m{y_A}^{k_A} (\bmod\ P))$ (i.e., the re-encryption
step), we use an algorithm similar to the ElGamal scheme:}
\begin{itemize}
\item \emph{Key generation: Let $x_B$ be the element uniformly
chosen from $Z_Q$, and $y_B\equiv g^{x_B}\bmod\ P$.}

\item \emph{Re-encryption: The re-encryption algorithm $E_R$, On
input $(P, Q, g, y_B)$ and a ciphertext $(y_1, y_2)=(g^{k_A}
(\bmod\ P), m{y_A}^{k_A} (\bmod\ P))$, uniformly selects an
element $k_B\in_rZ_Q$ and outputs
$$E_R((P, Q, g, y_B), y_1, y_2)=(y_1, g^{k_B} (\bmod\ P), y_2{y_B}^{k_B} (\bmod\ P))$$}

\end{itemize}
\end{itemize}

Note that since $(y_1, y_2)=(g^{k_A} (\bmod\ P), m{y_A}^{k_A}
(\bmod\ P))$, the ciphertext (after re-encryption) is
$$E_R((P, Q, g, y_B), y_1, y_2)=(g^{k_A} (\bmod\ P), g^{k_B} (\bmod\ P), m{y_A}^{k_A}{y_B}^{k_B} (\bmod\ P))$$
To simplify, let $(c_1, c_2, c_3)=(g^{k_A} (\bmod\ P), g^{k_B}
(\bmod\ P), m{y_A}^{k_A}{y_B}^{k_B} (\bmod\ P))$ and so \\
$E_R((P, Q, g, y_B), (y_1, y_2))=(c_1, c_2, c_3)$. Also, we use
$E_A$ and $E_B(E_A)$ to represent the encryption and re-encryption
processes respectively (with key $x_A$ and $x_B$).

The decryption is also similar to the ElGama scheme, but need to
decrypt twice, details follow:

\begin{itemize}
\item \emph{First round: The decryption algorithm, $D_B$, on input
$(P, Q, g, x_B)$ and a ciphertext $(c_1, c_2, c_3)$, outputs
$$D_B((P, g, x_B), (c_1,
c_2, c_3))=(c_1, c_3(c_2^{x_B})^{-1}(\bmod\ P))$$}
\end{itemize}
Now let us see what we obtain after this round: from $(c_1, c_2,
c_3)=(g^{k_A} (\bmod\ P), g^{k_B} (\bmod\ P),
m{y_A}^{k_A}{y_B}^{k_B} (\bmod\ P))$ we come up with $c_1=g^{k_A}
(\bmod\ P)$ and
$$c_3(c_2^{x_B})^{-1}(\bmod\ P)=my_A^{k_A}(\bmod\ P)$$
Thus we end up with $D_B((P, g, x_B), (c_1, c_2, c_3))=(y_1,
y_2)$, using ElGamal scheme we could decrypt the ciphertext $(y_1,
y_2)$:
\begin{itemize}
\item \emph{The decryption algorithm, $D_A$, on input $(P, Q, g,
x_A)$ and a ciphertext $(y_1, y_2)$, outputs $$D_A((P, g, x_A),
(y_1, y_2))=y_2(y_1^{x_A})^{-1} (\bmod\ P)(=m)$$}
\end{itemize}

In this paper, we directly present a theorem concerning the
security of the re-encryption scheme without proving it. For the
proof, we recommend readers to Ref.$\cite{elgamal security}$
\begin{theorem}
If the re-encryption scheme is not secure in the sense of
indistinguishability, then there exists a probabilistic
polynomial-time Turing Machine (p.p.t. TM) that solves the
decision Diffie-Hellman problem with overwhelming probability.
\end{theorem}

Furthermore, it is proved that breaking decision D-H problem is
almost as hard as computing discrete logarithms$\cite{DDH=DL}$,
while computing discrete logarithms is as hard as languages in NPC
unless the polynomial hierarchy (PH) collapses to the second
level$\cite{DL=NPC}$.

\section{Commutative-like encryption}

Commutative-like encryption is a new notion presented in this
paper, before giving the definition of commutative-like
encryption, let us first check one property of the above
re-encryption scheme.

In the decryption scheme, we decrypt the re-encrypted ciphertext
in a way corresponding to the order of encryption, however, we may
apply a different order, details follow:

\begin{itemize}
\item \emph{First round: The decryption algorithm, $D_A$, on input
$(P, Q, g, x_A)$ and a ciphertext $(c_1, c_2, c_3)$, outputs
$$D_A((P, g, x_A), (c_1,
c_2, c_3))=(c_2, c_3(c_1^{x_A})^{-1}(\bmod\ P))$$}
\end{itemize}
Now let us see what we obtain after this round: from $(c_1, c_2,
c_3)=(g^{k_A} (\bmod\ P), g^{k_B} (\bmod\ P),
m{y_A}^{k_A}{y_B}^{k_B} (\bmod\ P))$ we come up with $c_2=g^{k_B}
(\bmod\ P)$ and
$$c_3(c_1^{x_A})^{-1}(\bmod\ P)=my_B^{k_B}(\bmod\ P)$$
Thus we end up with $D_A((P, g, x_A), (c_1, c_2, c_3))=(y_1',
y_2')$, where $y_1'=g^{k_B}(\bmod\ P)$ and $y_2'=my_B^{k_B}(\bmod\
P)$ using ElGamal scheme we could decrypt the ciphertext $(y_1',
y_2')$:
\begin{itemize}
\item \emph{The decryption algorithm, $D_B$, on input $(P, Q, g,
x_B)$ and a ciphertext $(y_1', y_2')$, outputs $$D_B((P, g, x_B),
(y_1', y_2'))=y_2(y_1^{x_B})^{-1} (\bmod\ P)$$}
\end{itemize}

Clearly, in both decryption schemes, we have the plaintext at the
last step. This suggests a ``commutative-like'' characterization:
the result of decryption does not relies on the order of
decryptions, more specifically, in the scheme, let $m$ be the
plaintext and $(c_1, c_2, c_3)$ be the ciphertext, we have
$$D_A(D_B(c_1, c_2, c_3))=D_B(D_A(c_1, c_2, c_3))=m$$
or equivalently, we have
$$D_B(D_A(E_B(E_A(m))))=m$$
Largely due to the probabilistic nature, this encryption cannot be
termed as commutative encryption, since the each ciphertext of the
same plaintext would be different in different time with
overwhelming probability, or say, $(c_1, c_2, c_3)=E_B(E_A(m))$ is
not fixed(in fact, the ciphertext is same unless the randomly
chosen variables $k_1, k_2$ are fixed).

DEFINITION 4\ \ (Commutative-like Encryption). \emph{A
commutative-like encryption $\mathcal{F}$ is a computable (in
polynomial time) function $f$ : Key\ $\mathcal{F}\times$\ Dom\
$\mathcal{F}\rightarrow$\ Ran\ $\mathcal{F}$, defined on finite
computable domains, that satisfies all properties listed below. }

\emph{ 1. Commutative-like. For all $e, e'\in$\ Key\ $\mathcal{F}$
we have
$$f^{-1}_{e'}\circ f^{-1}_e\circ f_{e'}\circ f_e=I$$}

\emph{2. The inverse $f^{-1}_e$ is is a deterministic process
(i.e., every ciphertext maps only one plaintext, while a plaintext
might map many ciphertext) and is also computable in
polynomial-time given $e$.}

\emph{3. The distribution of $\langle x, f_e(x), y, f_e(y)\rangle$
is indistinguishable from the distribution of $\langle x, f_e(x),
y, z\rangle$, where $x, y\in_r$\ Dom\ $\mathcal{F}$, $z\in_r$ Ran
$\mathcal{F}$ and $e\in_r$ \ Key\ $\mathcal{F}$.}

Informally, Property 1 says that when we compositely encrypt with
two different keys, the result is the same irrespective of the
order of decryption. Property 2 says that given an encrypted value
$f_e(x)$ and the encryption key $e$, we can find $x$ in polynomial
time, and there is only one such $x$. Property 3 says that given a
value $x$ and its encryption $f_e(x)$ (but not the key $e$), for a
new value $y$, we cannot distinguish between $f_e(y)$ and a random
value $z$ in polynomial time. Thus we can neither encrypt $y$ nor
decrypt $f_e(y)$ in polynomial time. Note that this property holds
only if $x$ is a random value from Dom $\mathcal{F}$, i.e., the
adversary does not control the choice of $x$.

Now let us see how the encryption scheme fits the required
properties. Obviously, the first and second properties comes
directly from the algorithms, now we check the third property.
Note that if $\langle x, f_e(x), y, f_e(y)\rangle=\langle m_1,
g^{k_A}, m_1g^{k_Ax}, m_2, g^{k_B}, m_2g^{k_Bx}\rangle$ (where
$(\bmod\ P)$ is neglected) is distinguishable from $\langle m_1,
g^{k_A}, m_1g^{k_Ax}, m_2, z_1, z_2\rangle$
$(z_1,z_2\in$Ran$\mathcal{F})$, then $\langle g^{k_A}, g^{k_Ax},
g^{k_B}, g^{k_Bx}\rangle$ is distinguishable from the distribution
of $\langle g^{k_A}, g^{k_Ax}, g^{k_B}, z\rangle$ where $z\in_r
Z_Q$. the Decisional Diffie-Hellman hypothesis (DDH) claims that
for any generating ($\neq 1$) element $g$, the distribution of
$\langle g^a, g^b, g^{ab}\rangle$ is indistinguishable from the
distribution of $\langle g^a, g^b, g^c\rangle$. A 3-tuple $\langle
g^a, g^b, z\rangle$ from the DDH can be reduced to our 4-tuple
$\langle g^{k_A}, g^{k_Ax}, g^{k_B}, z\rangle$ by taking
$d\in$Key$\mathcal{F}$ and making tuple $\langle g^d, (g^a)^d,
g^b, z\rangle$. Now $a$ plays the role of $x$, $g^d$ of $g^{k_A}$,
and $g^b$ of $g^{k_B}$; we test whether $g^{ab}$ or is random.
Thus, given DDH, $\langle g^{k_A}, g^{k_Ax}, g^{k_B},
g^{k_Bx}\rangle$ and $\langle g^{k_A}, g^{k_Ax}, g^{k_B},
z\rangle$ are also indistinguishable, which contradicts our
assumption.

\section{Application Instance}
Readers might wonder the real application of commutative-like
encryption, and here we propose one possible application in
oblivious transfer. Oblivious Transfer refers to a kind of
two-party protocols where at the beginning of the protocol one
party, the sender, has an input, and at the end of the protocol
the other party, the receiver, learns some information about this
input in a way that does not allow the sender to figure out what
it has learned. Oblivious transfer is a fundamental primitive in
the design and analysis of cryptographic protocols$\cite{ot1,
ot2}$. Our scheme is a 1-out-of-$n$ oblivious transfer: the sender
has $n$ secrets $m_1, m_2, \ldots, m_n$ and is willing to disclose
exactly one of them to the receiver at its choice.

Now let us see how our protocol proceeds:

\begin{itemize}
\item The sender encrypts every item using its key $x_A$ and gets
$E_{x_A}(m_1), E_{x_A}(m_2), \cdots, E_{x_A}(m_n)$. Then it
reveals them to the receiver.
\end{itemize}

\begin{itemize}
\item On receiving the ciphertexts, the receiver chooses exactly
one of them, say, $E_{x_A}(m_i)(1\leq i\leq n)$, and encrypts it
to obtain $E_{x_B}(E_{x_A}(m_i))$ and tells it to the sender.
\end{itemize}

\begin{itemize}
\item The sender decrypts it, gets
$D_{x_A}(E_{x_B}(E_{x_A}(m_i)))$ and sends it to the receiver.
\end{itemize}

\begin{itemize}
\item The receiver obtains $m_i$ by calculating
$D_{x_B}(D_{x_A}(E_{x_B}(E_{x_A}(m_i))))$.
\end{itemize}

Instead of a formal proof, we explain how the protocol achieves
its goal: according to the performance of commutative-like
encryption, the receiver can get its desired message after
interaction with the sender, i.e.,
$D_{x_B}(D_{x_A}(E_{x_B}(E_{x_A}(m_i))))=D_{x_A}(D_{x_B}(E_{x_B}(E_{x_A}(m_i))))=m_i$,
thus the protocol is correct. Furthermore, the receiver receives
nothing other than $m_i$: it can hardly deduce anything from the
ciphertexts $E_{x_A}(m_i)(1\leq i\leq n)$. As for the privacy of
the receiver, the sender does not know the receiver's choice $i$:
it does not suggest $m_i$ from $E_{x_B}(E_{x_A}(m_i))$.

It should noted that by trivially perform the protocol $m$ times,
we would obtain an $m$-out-of-$n$ oblivious transfer protocol.

\section{Conclusion}
In this paper, we define the notion of commutative-like
encryption, which is a useful variation of commutative encryption.
As an example, it is showed that the ElGamal scheme could be such
a commutative-like scheme. Also, we discussed one possible
application of commutative-like encryption.

\end{document}